\documentclass[pra,twocolumn,showpacs,preprintnumbers,amsmath,amssymb]{revtex4}

\usepackage{graphicx}
\usepackage{dcolumn}
\usepackage{bm}

\begin{document}

\title{Attacking quantum key distribution with single-photon two-qubit quantum logic}

\author{Jeffrey H. Shapiro}
\email[Electronic address: ]{jhs@mit.edu}
\author{Franco N. C. Wong}
\affiliation{Massachusetts Institute of Technology, Research Laboratory of Electronics, Cambridge, Massachusetts 02139 USA}

\date{\today}

\begin{abstract}
The Fuchs-Peres-Brandt (FPB) probe realizes the most powerful individual attack on Bennett-Brassard 1984  quantum key distribution (BB84 QKD) by means of a single controlled-NOT (CNOT) gate.  This paper describes a complete physical simulation of the FPB-probe attack on polarization-based BB84 QKD using a deterministic CNOT constructed from single-photon two-qubit quantum logic.  Adding polarization-preserving quantum nondemolition measurements of photon number to this configuration converts the physical simulation into a true deterministic realization of the FPB attack.  
\end{abstract}

\maketitle

\section{Introduction}

Bennett-Brassard 1984 quantum key distribution (BB84 QKD) using single-photon polarization states works as follows \cite{BB84}.  In each time interval allotted for a bit, Alice transmits a single photon in a randomly selected polarization, chosen from horizontal ($H$), vertical ($V$), $+$45$^\circ$, or 
$-$45$^\circ$, while Bob randomly chooses to detect photons in either the $H$/$V$ or $\pm$45$^\circ$ bases.  Bob discloses to Alice the sequence of bit intervals and associated measurement bases for which he has detections.  Alice then informs Bob which detections occurred in bases coincident with the ones that she used.  These are the \em sift\/\rm\ events, i.e., bit intervals in which Bob has a detection \em and\/\rm\
his count has occurred in the same basis that Alice used.  An \em error\/\rm\  event is a sift event in
which Bob decodes the incorrect bit value.   Alice and Bob employ a  prescribed set of operations to identify errors in their sifted bits, correct these errors, and apply sufficient privacy amplification to deny useful key information to any potential eavesdropper (Eve).  At the end of the full QKD procedure, Alice and Bob have a shared one-time pad with which they can communicate in complete security.

In long-distance QKD systems, most of Alice's photons will go undetected, owing to propagation loss and detector inefficiencies.  Dark counts and, for atmospheric QKD systems, background counts can cause error events in these systems, as can intrusion by Eve.  Employing an attenuated laser source, in lieu of a true single-photon source, further reduces QKD performance as such sources are typically run at less than one photon on average per bit interval, and the occurrence of multi-photon events, although rare at low average photon number, opens up additional vulnerability.  Security proofs  have been published for ideal BB84  \cite{security1}, as have security analyses that incorporate a variety of non-idealities \cite{security2}.  Our attention, however, will be directed toward attacking BB84 QKD, as to our knowledge no such experiments have been performed, although a variety of potentially practical approaches have been discussed \cite{attacks}.  Our particular objective will be to show that current technology permits physical simulation of the Fuchs-Peres-Brandt (FPB) probe \cite{FPB}, i.e., the most powerful individual attack on single-photon BB84, and that developments underway in quantum nondemolition (QND) detection may soon turn this physical simulation into a full implementation of the attack.  Thus we believe it is of interest to construct the physical simulation and put BB84's security to the test:  how much information can Eve really derive about the key that Alice and Bob have distilled while keeping Alice and Bob oblivious to her presence.  

The remainder of this paper is organized as follows.  In Sec.~II we review the FPB probe and its theoretical performance.  In Sec.~III we describe a complete physical simulation of this probe constructed from single-photon two-qubit (SPTQ) quantum logic.  We conclude, in Sec.~IV, by showing how the addition of polarization-preserving QND measurements of photon number can convert this physical simulation into a true deterministic realization of the FPB attack on polarization-based BB84.   

\section{The Fuchs-Peres-Brandt Probe}
In an individual attack on single-photon BB84 QKD, Eve probes Alice's photons one at a time.   In a collective attack, Eve's measurements probe groups of Alice's photons.  Less is known about collective attacks \cite{collective}, so we will limit our consideration to individual attacks.  Fuchs and Peres \cite{FP} described the most  general way in which an individual attack could be performed.  Eve supplies a probe photon and lets it interact with Alice's photon in a unitary manner.   Eve then sends Alice's photon  to Bob, and performs a probability operator-valued measurement (POVM) on the probe photon she has retained.  Slutsky {\em et al.} \cite{Slutsky} demonstrated that the Fuchs-Peres construct---with the appropriate choice of probe state, interaction, and measurement---affords Eve the maximum amount of R\'{e}nyi information about the error-free sifted bits that Bob receives for a given level of disturbance, i.e., for a given probability that a sifted bit will be received in error.  Brandt \cite{FPB} extended the Slutsky {\em et al.} treatment by showing that the optimal probe could be realized with a single CNOT gate.  Figure~1 shows an abstract diagram of the resulting Fuchs-Peres-Brandt probe.  In what follows we give a brief review of its structure and performance---see \cite{FPB} for a more detailed treatment---where, for simplicity, we assume ideal conditions in which Alice transmits a single photon per bit interval, there is no propagation loss and no extraneous (background) light collection, and both Eve and Bob have unity quantum efficiency photodetectors with no dark counts.  These ideal conditions imply there will not be any errors on sifted bits in the absence of eavesdropping; the case of more realistic conditions will be discussed briefly in Sec.~IV.
\begin{figure}[h]
\includegraphics[width = 3.4in]{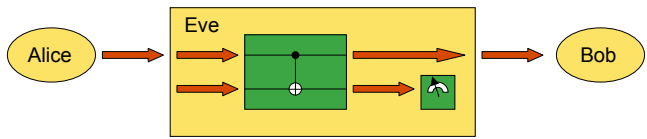}
\caption{(Color online) Block diagram of the Fuchs-Peres-Brandt probe for attacking BB84 QKD.}
\end{figure}

In each bit interval Alice transmits, at random, a single photon in one of the four BB84 polarization states.  Eve uses this photon as the control-qubit input to a CNOT gate whose computational basis---relative to the BB84 polarization states---is shown in Fig.~2, namely 
\begin{eqnarray}
|0\rangle &\equiv& \cos(\pi/8)|H\rangle + \sin(\pi/8)|V\rangle \\ 
|1\rangle &\equiv& -\sin(\pi/8)|H\rangle + \cos(\pi/8)|V\rangle,
\end{eqnarray} 
in terms of the $H/V$ basis.  Eve supplies her own probe photon, as the target-qubit input to this CNOT gate, in the state
\begin{equation}
|T_{\rm in}\rangle \equiv C|+\rangle + S|-\rangle,
\label{probeinput}
\end{equation}
where $C = \sqrt{1-2P_E}$, $S = \sqrt{2P_E}$, $|\pm\rangle = (|0\rangle \pm |1\rangle)/\sqrt{2}$, and $0\le P_E\le 1/2$ will turn out to be the error probability that Eve's probe creates on Bob's sifted bits \cite{footnote1}.  So, as $P_E$ increases from 0 to 1/2, $|T_{\rm in}\rangle$ goes from $|+\rangle$ to $|-\rangle$.  The (unnormalized) output states that may occur for this target qubit are
\begin{eqnarray}
|T_\pm\rangle &\equiv& C|+\rangle \pm \frac{S}{\sqrt{2}}|-\rangle \\ 
|T_E\rangle &\equiv& \frac{S}{\sqrt{2}}|-\rangle.
\end{eqnarray}
\begin{figure}
\includegraphics[width = 2.4in]{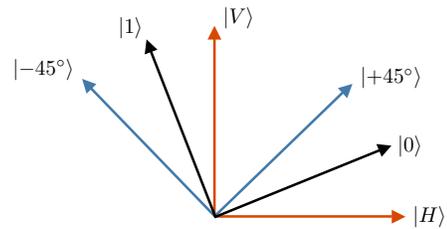}
\caption{(Color online) Computational basis for Eve's CNOT gate referenced to the BB84 polarization states.}
\end{figure}

Here is how the FPB probe works.  When Alice uses the $H/V$ basis for her photon transmission, Eve's CNOT gate effects the following transformation,
\begin{eqnarray}
|H\rangle|T_{\rm in}\rangle &\longrightarrow&
|H\rangle|T_-\rangle + |V\rangle|T_E\rangle \label{Hin_out} \\ 
|V\rangle|T_{\rm in}\rangle &\longrightarrow&
|V\rangle|T_+\rangle +|H\rangle|T_E\rangle,\label{Vin_out} 
\end{eqnarray}
where the kets on the left-hand side denote the Alice\,$\otimes$\,Eve state of the control and target qubits at the CNOT's input and the kets on the right-hand side denote the Bob\,$\otimes$\,Eve state of the control and target qubits at the CNOT's output.  
Similarly, when Alice uses the $\pm 45^\circ$ basis, Eve's CNOT gate has the following behavior,
\begin{eqnarray}
|\mbox{$+$}45^\circ\rangle|T_{\rm in}\rangle &\longrightarrow&
|\mbox{$+$}45^\circ\rangle|T_+\rangle + |\mbox{$-$}45^\circ\rangle|T_E\rangle 
\label{plus_in_out}\\ 
|\mbox{$-$}45^\circ\rangle|T_{\rm in}\rangle &\longrightarrow&
|\mbox{$-$}45^\circ\rangle|T_-\rangle +|\mbox{$+$}45^\circ\rangle|T_E\rangle.
\label{minus_in_out}
\end{eqnarray}
Suppose that Bob measures in the basis that Alice has employed \em and\/\rm\ his outcome matches what Alice sent.  Then Eve can learn their shared  bit value, once Bob discloses his measurement basis, by distinguishing between the $|T_+\rangle$ and $|T_-\rangle$ output states for her target qubit.  Of course, this knowledge comes at a cost:  Eve has caused an error event whenever Alice and Bob choose a common basis and her target qubit's output state is $|T_E\rangle$.  To maximize the information she derives from this intrusion, Eve applies the minimum error probability receiver for distinguishing between
the single-photon polarization states $|T_+\rangle$ and $|T_-\rangle$.  This is a projective measurement onto the polarization basis $\{|d_+\rangle,|d_-\rangle\}$, shown in Fig.~3 and given by
\begin{eqnarray}
|d_+\rangle &=& \frac{|+\rangle + |-\rangle}{\sqrt{2}} = |0\rangle \\ 
|d_-\rangle &=& \frac{|+\rangle - |-\rangle}{\sqrt{2}} = |1\rangle. 
\end{eqnarray}
\begin{figure}[h]
\includegraphics[width = 2.2in]{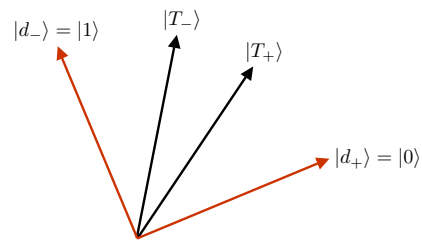}
\caption{(Color online) Measurement basis for Eve's minimum-error-probability discrimination between $|T_+\rangle$ and $|T_-\rangle$.}
\end{figure}

Two straightforward calculations will now complete our review of the FPB probe.  First, we find the error probability that is created by Eve's presence.  Suppose Alice and Bob use the $H/V$ basis and Alice has sent $|H\rangle$.  Alice and Bob will incur an error if the control\,$\otimes$\,target output from Eve's CNOT gate is $|V\rangle|T_E\rangle$.  The probability that this occurs is $\langle T_E| T_E\rangle = S^2/2 = P_E$.  The same conditional error probability ensues for the other three error events, e.g., when Alice and Bob use the $\pm 45^\circ$ basis, Alice sends $|$$+45^\circ\rangle$, and the CNOT
output is $|$$-45^\circ\rangle|T_E\rangle$.  It follows that the unconditional error probability incurred by Alice and Bob on their sift events is $P_E$.  

Now we shall determine the R\'{e}nyi information that Eve derives about the sift events for which Alice and Bob do not suffer errors.    Let $B = \{0,1\}$ and $E = \{0,1\}$ denote the ensembles of possible bit values that Bob and Eve receive on a sift event in which Bob's bit value agrees with Alice's.  The R\'{e}nyi information (in bits) that Eve learns about each Alice/Bob error-free sift event is
\begin{eqnarray}
I_R &\equiv&
-\log_2\!\left(\sum_{b= 0}^1P^2(b)\right) \nonumber \\
&+&\sum_{e = 0}^1P(e)\log_2\!\left(\sum_{b = 0}^1
P^2(b\mid e)\right),
\end{eqnarray}
where $\{P(b), P(e)\}$ are the prior probabilities for Bob's and Eve's bit values, and $P(b\mid e)$ is the conditional probability for Bob's bit value to be $b$ given that Eve's is $e$.  Alice's bits are equally likely to be 0 or 1, and Eve's conditional error probabilities satisfy \cite{Helstrom}
\begin{eqnarray}
\lefteqn{P(e = 1\mid b = 0) = P(e = 0\mid b = 1)} \\ 
&=& \frac{1}{2}\!\left(1 - \sqrt{1 - \frac{|\langle T_+|T_-\rangle|^2}{\langle T_+|T_+\rangle \langle T_-|T_-\rangle}}\right) \\
&=& \frac{1}{2}\!\left(1- \frac{\sqrt{4P_E(1-2P_E)}}{1-P_E}\right).
\end{eqnarray}
These results imply that $b$ is also equally likely to be 0 or 1, and that $P(b\mid e) = P(e\mid b)$, whence
\begin{equation}
I_R = \log_2\!\left(1 + \frac{4P_E(1-2P_E)}{(1-P_E)^2}\right),
\end{equation}
which we have plotted in Fig.~4.
\begin{figure}[h]
\vspace*{.2in}
\includegraphics[width = 2in]{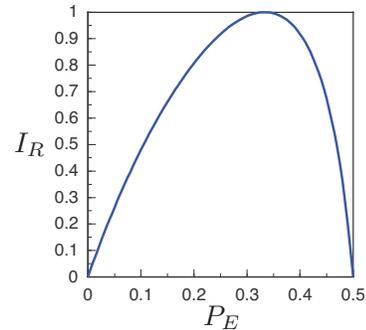}
\caption{(Color online) Eve's R\'{e}nyi information about Bob's error-free sifted bits as a function of the error probability that her eavesdropping creates.}
\end{figure}

Figure~4 reveals several noteworthy performance points for the FPB probe.  The $I_R = 0, P_E = 0$ point in this figure corresponds to Eve's operating her CNOT gate with $|T_{\rm in}\rangle = |+\rangle$ for its target qubit input.  It is well known that such an input is unaffected by and does not affect the control qubit.  Thus Bob suffers no errors but Eve gets no R\'{e}nyi information.  The $I_R = 1, P_E = 1/3$ point in this figure corresponds to Eve's operating her CNOT gate with $|T_{\rm in}\rangle = \sqrt{1/3}|+\rangle + \sqrt{2/3}|-\rangle$, which leads to $|T_\pm\rangle \propto |d_\pm\rangle$.  In this case Eve's Fig.~3 receiver makes no errors, so she obtains the maximum (1 bit) R\'{e}nyi information about each of Bob's error-free bits.  The $I_R = 0, P_E = 1/2$ point in this figure corresponds to Eve's operating her CNOT gate with $|T_{\rm in}\rangle = |-\rangle$, which gives $|T_+\rangle = |T_-\rangle = |T_E\rangle = \sqrt{1/2}|-\rangle$.  Here it is clear that Eve gains no information about Bob's error-free bits, but his error probability is 1/2 because of the action of the $|-\rangle$ target qubit on the control qubit.  

\section{Physical Simulation in SPTQ Logic}
In single-photon two-qubit quantum logic, each photon encodes two independently controllable qubits \cite{SPTQ1}.  One of these is the familiar polarization qubit, with basis $\{|H\rangle,|V\rangle\}$.  The other we shall term the momentum qubit---because our physical simulation of the FPB probe will rely on the polarization-momentum hyperentangled photon pairs produced by type-II phase matched spontaneous parametric downconversion (SPDC)---although in the collimated configuration in which SPTQ is implemented its basis states are single-photon kets for right and left beam positions (spatial modes), denoted $\{|R\rangle, |L\rangle\}$.  Unlike the gates proposed for linear optics quantum computing \cite{KLM}, which are scalable but non-deterministic, SPTQ quantum logic is deterministic but not scalable.  Nevertheless, SPTQ quantum logic suffices for a complete physical simulation of polarization-based BB84 being attacked with the FPB probe, as we shall show.  Before doing so, however, we need to comment on the gates that have been demonstrated in SPTQ logic.

It is well known that single qubit rotations and CNOT gates form a universal set for quantum computation.  In SPTQ quantum logic, polarization-qubit rotations are easily accomplished with wave plates, just as is done in linear optics quantum computing.  Momentum-qubit rotations are realized by first performing a SWAP operation, to exchange the polarization and momentum qubits, then rotating the polarization qubit, and finally performing another SWAP.  The SWAP operation is a cascade of three CNOTs, as shown in Fig.~5.  For its implementation in SPTQ quantum logic the left and right CNOTs in Fig.~5 are momentum-controlled NOT gates (M-CNOTs) and the middle CNOT is a polarization-controlled NOT gate (P-CNOT).   (An M-CNOT uses the momentum qubit of a single photon to perform the controlled-NOT operation on the polarization qubit of that same photon, and vice versa for the P-CNOT gate.)  Experimental demonstrations of deterministic M-CNOT, P-CNOT, and SWAP gates are reported in \cite{SPTQ1,SPTQ2}.  
\begin{figure}[h]
\includegraphics[width = 2in]{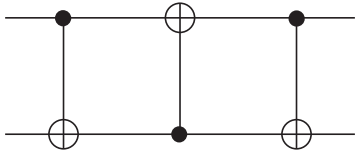}
\caption{Quantum circuit diagram for a SWAP gate realized as a cascade of three CNOTs.  In SPTQ quantum logic the upper rail is the momentum qubit and the lower rail is the polarization qubit of the same photon.}
\end{figure}

Figure~6 shows a physical simulation of polarization-based BB84 under FPB attack when Alice has a single-photon source and Bob employs active basis selection; Fig.~7 shows the modification needed to accommodate Bob's using passive basis selection.    In either case, Alice uses a polarizing beam splitter and an electro-optic modulator, as a controllable half-wave plate (HWP), to set the randomly-selected BB84 polarization state for each photon she transmits.  Moreover, she employs a single spatial mode, which we assume coincides with the $R$ beam position in Eve's apparatus.  Eve then begins her attack by imposing the probe state $|T_{\rm in}\rangle$ on the momentum qubit.  She does this by applying a SWAP gate, to exchange the momentum and polarization qubits of Alice's photon, rotating the resulting polarization qubit (with the HWP in Fig.~6) to the $|T_{\rm in}\rangle$ state, and then using another SWAP to switch this state into the momentum qubit.  This procedure leaves Alice's BB84 polarization state unaffected, although her photon, which will ultimately propagate on to Bob, is no longer in a single spatial mode.  Eve completes the first stage of her attack by sending Alice's photon through a P-CNOT gate, which will accomplish the state transformations given in Eqs.~(\ref{Hin_out})--(\ref{minus_in_out}), and then routing it to Bob.  If Bob employs active basis selection (Fig.~6), then in each bit interval he will use an electro-optic modulator---as a controllable HWP---plus a polarizing beam splitter to set the randomly-selected polarization basis for his measurement.  The functioning of this basis-selection setup is unaffected by Alice's photon no longer being in a single spatial mode.  The reason that we call Fig.~6 a physical simulation, rather than a true attack, lies in the measurement box.  Here, Eve has invaded Bob's turf, and inserted SWAP gates, half-wave plates, polarizing beam splitters, and additional photodetectors, so that she can forward to Bob measurement results corresponding to photon counting on the polarization basis that he has selected while she retains the photon counting results corresponding to her $\{|d_+\rangle, |d_-\rangle\}$ measurement.  Clearly Bob would never knowingly permit Eve to intrude into his receiver box in this manner.  Moreover, if Eve could do so, she would not bother with an FPB probe as she could directly observe Bob's bit values.  
\begin{figure}[h]
\vspace*{.1in}
\includegraphics[width = 3.4in]{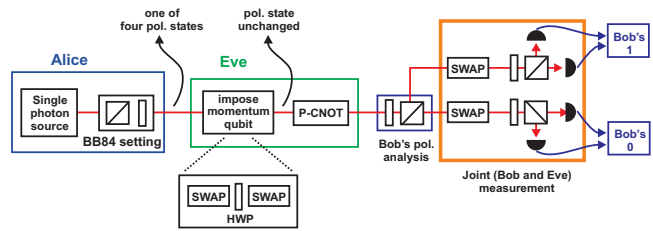}
\caption{(Color online) Physical simulation of polarization-based BB84 QKD and the FPB-probe attack.}
\end{figure}
\begin{figure}[h]
\vspace*{.1in}
\includegraphics[width = 3.4in]{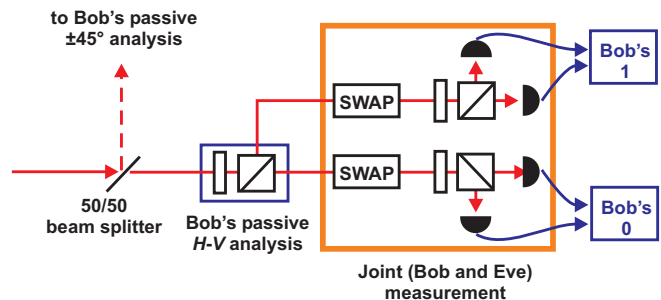}
\caption{(Color online) Modification of Fig.~6 to accommodate Bob's using passive basis selection.}
\end{figure}

If Bob employs passive basis selection (Fig.~7), then he uses a 50/50 beam splitter followed by static-HWP analysis in the $H$-$V$ and $\pm 45^\circ$ bases, with only the former being explicitly shown in Fig.~7.  The rest of Eve's attack mimics what was seen in Fig.~6, i.e., she gets inside Bob's measurement boxes with SWAP gates, half-wave plates, and additional detectors so that she can perform her probe measurement while providing Bob with his BB84 polarization-measurement data.  Because the Fig.~7 arrangement requires that twice as many SWAP gates, twice as many half-wave plates, and twice as many single-photon detectors be inserted into Bob's receiver system, as compared to what is needed in the Fig.~6 setup, we shall limit the rest of our discussion to the case of active basis selection as it leads to a more parsimonious physical simulation of the Fuchs-Peres-Brandt attack.  We recognize, of course, that the decision to use active basis selection is Bob's to make, not Eve's.  More importantly, however, in Sec.~IV we will show how the availability of polarization-preserving QND photon-number measurements can be used to turn Fig.~6 into a true, deterministic implementation of the FPB attack. The same conversion can be accomplished for passive basis selection.  Before turning to the true-attack implementation, let us flesh out some details of the measurement box in Fig.~6 and show how SPDC can be used, in lieu of the single-photon source, to perform this physical simulation.

Let $|\psi_{\rm out}\rangle$ denote the polarization\,$\otimes$\,momentum state at the output of Eve's P-CNOT gate in Fig.~6.  Bob's polarization analysis box splits this state, according to the basis he has chosen, so that one basis state goes to the upper branch of the measurement box while the other goes to the lower branch of that box.  This polarization sorting does nothing to the momentum qubit, so the SWAP gates, half-wave plates, and polarizing beam splitters that Eve has inserted into the measurement box accomplish her $\{|d_+\rangle, |d_-\rangle\}$ projective measurement, i.e., the horizontal paths into photodetectors in Fig.~6 are projecting the momentum qubit of $|\psi_{\rm out}\rangle$ onto $|d_-\rangle$ and the vertical paths into photodetectors in Fig.~6 are projecting this state onto 
$|d_+\rangle$.   Eve records the combined results of the two $|d_+\rangle$ versus $|d_-\rangle$ detections, whereas Bob, who only sees the combined photodetections for the upper and lower branches entering the measurement box, gets his BB84 polarization data.  Bob's data is impaired, of course, by the effect of Eve's P-CNOT.  

Single-photon on-demand sources are now under development at several institutions \cite{single}, and their use in BB84 QKD has been demonstrated \cite{singleBB84}.  At present, however, it is much more practical to use SPDC as a heralded source of single photons \cite{herald}.  In SPDC, signal and idler photons are emitted in pairs, thus detection of the signal photon heralds the presence of the idler photon.  Moreover, with appropriate configurations \cite{bidirectional}, SPDC will produce photons that are simultaneously entangled in polarization and in momentum.  This hyperentanglement leads us to propose the Fig.~8 configuration for physically simulating the FPB-probe attack on BB84.  Here, a pump laser drives SPDC in a type-II phase matched $\chi^{(2)}$ crystal, such as periodically-poled potassium titanyl phosphate (PPKTP), producing pairs of orthogonally-polarized, frequency-degenerate photons that are entangled in both polarization and momentum.  The first polarizing beam splitter transmits a horizontally-polarized photon and reflects a vertically-polarized photon while preserving their momentum entanglement.  Eve uses a SWAP gate and (half-wave plate plus polarizing beam splitter) polarization rotation so that her photodetector's clicking will, by virtue of the momentum entanglement, herald the setting of the desired $|T_{\rm in}\rangle$ momentum-qubit state on the horizontally-polarized photon emerging from the first polarizing beam splitter.  Alice's electronically controllable half-wave plate sets the BB84 polarization qubit on this photon, and the rest of the Fig.~8 configuration is identical to that shown and explained in Fig.~6.    Inasmuch as the SPDC source and SPTQ gates needed to realize the Fig.~8 setup have been demonstrated, we propose that such an experiment be performed.  Simultaneous recording of Alice's polarization choices, Bob's polarization measurements and Eve's $|d_+\rangle$ versus $|d_-\rangle$ results can then be processed through the BB84 protocol stack to study the degree to which the security proofs and eavesdropping analyses stand up to experimental scrutiny.  
\begin{figure}[h]
\vspace*{.1in}
\includegraphics[width = 3.4in]{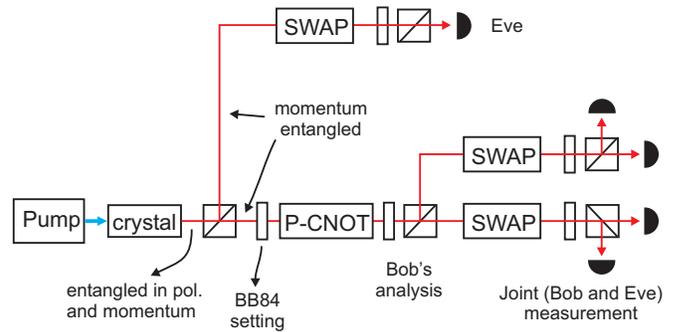}
\caption{(Color online) Proposed configuration for a complete physical simulation of the FPB attack on BB84 that is based on hyperentangled photon pairs from type-II phase matched SPDC and gates built from SPTQ quantum logic.}
\end{figure}

\section{The Complete Attack}
Although the FPB attack's physical simulation, as described in the preceding section, is both experimentally feasible and technically informative, any vulnerabilities it might reveal would only be of academic interest were there no practical means to turn it into a true deterministic implementation in which Eve did \em not\/\rm\ need to invade Bob's receiver.  Quantum nondemolition measurement technology provides the key to creating this complete attack.  As shown in the appendix, it is possible, in principle, to use cross-phase modulation between a strong coherent-state probe beam and an arbitrarily polarized signal beam to make a QND measurement of the signal beam's total photon number while preserving its polarization state.  Cross-phase modulation QND measurement of photon number has long been a topic of interest in quantum optics \cite{Imoto}, and recent theory has shown that it provides an excellent new route to photonic quantum computation \cite{Nemoto}.  Thus it is not unwarranted to presume that polarization-preserving QND measurement of total photon number may be developed.  With such technology in hand, the FPB-probe attack shown in Fig.~9 becomes viable.  Here, Eve imposes a momentum qubit on Alice's polarization-encoded photon and performs a P-CNOT operation exactly as discussed in conjunction with Figs.~6 and 8.  Now, however, Eve uses a SWAP-gate  half-wave plate combination so that the $|d_+\rangle$ and $|d_-\rangle$ momentum qubit states emerging from her P-CNOT become $|V\rangle$ and $|H\rangle$ states entering the polarizing beam splitter that follows the half-wave plate.  This beam splitter routes these polarizations into its transmitted and reflected output ports, respectively, where, in each arm, Eve employs a SWAP gate, a polarization-preserving QND measurement of total photon number, and another SWAP gate.  The first of these SWAPs returns Alice's BB84 qubit to polarization, so that a click on Eve's polarization-preserving QND apparatus completes  her $\{|d_+\rangle, |d_-\rangle\}$ measurement without further scrambling Alice's BB84 qubit beyond what has already occurred in Eve's P-CNOT gate.  The SWAP gates that follow the QND boxes then restore definite ($V$ and $H$) polarizations to the light in the upper and lower branches so that they may be recombined on a polarizing beam splitter.  The SWAP gate that follows this recombination then returns the BB84 qubit riding on Alice's photon to polarization for transmission to and measurement by Bob.  This photon is no longer in the single spatial mode emitted by Alice's transmitter, hence Bob could use spatial-mode discrimination to infer the presence of Eve, regardless of the $P_E$ value she had chosen to impose.  Eve, however, can preclude that possibility.  Because the result of her $\{|d_+\rangle,|d_-\rangle\}$ measurement tells her the value of the momentum qubit on the photon being sent to Bob, she can employ an additional stage of qubit rotation to restore this momentum qubit to the $|R\rangle$ state corresponding to Alice's transmission.  Also, should Alice try to defeat Eve's FPB probe by augmenting her BB84 polarization qubit with a randomly-chosen momentum qubit, Eve can use a QND measurement setup like that shown in Fig.~9 to collapse the value of that momentum qubit to $|R\rangle$ or $|L\rangle$, and then rotate that momentum qubit into the $|R\rangle$-state spatial mode before applying the FPB-probe attack.  At the conclusion of her attack, she can then randomize the momentum qubit on the photon that will be routed on to Bob without further impact---beyond that imposed by her P-CNOT gate---on that photon's polarization qubit.  So, unless Alice and Bob generalize their polarization-based BB84 protocol to include cooperative examination of the momentum qubit, Alice's randomization of that qubit will neither affect Eve's FPB attack, nor provide Alice and Bob with any additional evidence, beyond that obtained from the occurrence of errors on sifted bits, of Eve's presence.  
\begin{figure}[h]
\vspace*{.1in}
\includegraphics[width = 3.4in]{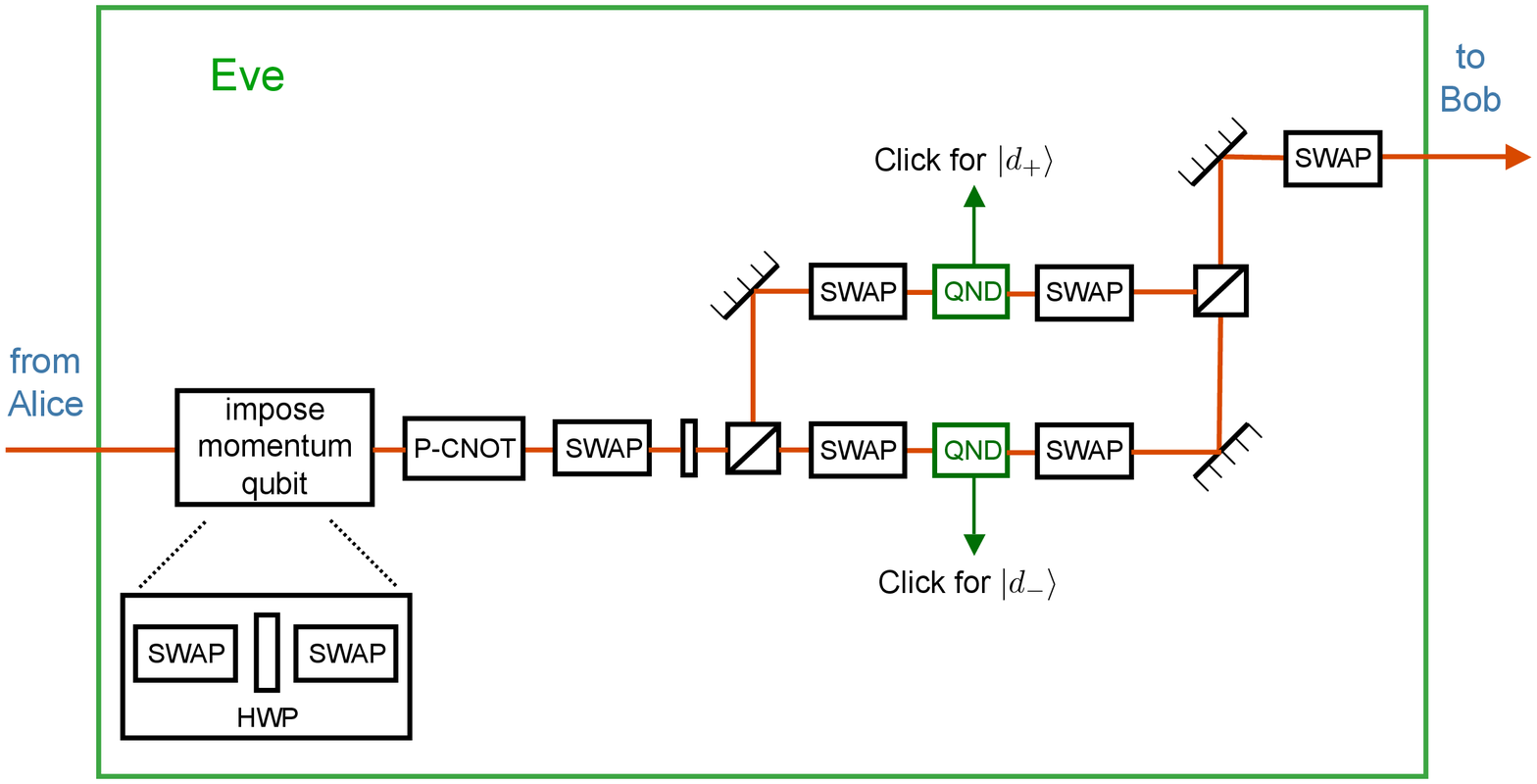}
\caption{(Color online) Deterministic FPB-probe attack on polarization-based BB84 that is realized with polarization-preserving QND measurements and SPTQ quantum logic.}
\end{figure}

Some concluding remarks are now in order.  We have shown that a physical simulation of the Fuchs-Peres-Brandt attack on polarization-based BB84 is feasible with currently available technology, and we have argued that the development of polarization-preserving QND technology for measuring total photon number will permit mounting of a true deterministic FBP-probe attack.  Our analysis has presumed ideal conditions in which Alice employs a single-photon source, there is no propagation loss and no extraneous (background) light collection, and both Eve and Bob have unity quantum efficiency photodetectors with no dark counts.  Because current QKD systems typically employ attenuated laser sources, and suffer from propagation loss, photodetector inefficiencies, and extraneous counts, it behooves us to at least comment on how such non-idealities could impact the FPB probe we have described.  

The use of an attenuated laser source poses no problem for the configurations shown in Figs.~6--9.  This is because the single-qubit rotations and the CNOT gates of SPTQ quantum logic effect the same transformations on coherent states as they do on single-photon states.  For example, the same half-wave plate setting that rotates the single-photon $|H\rangle$ qubit into the single-photon $|V\rangle$ qubit  will transform the horizontally-polarized coherent state $|\alpha\rangle_H$ into the vertically-polarized coherent state $|\alpha\rangle_V$.  Likewise, the SPTQ P-CNOT gate that transforms a single photon carrying polarization ($|H\rangle = |0\rangle, |V\rangle = |1\rangle$) and momentum ($|R\rangle = |0\rangle, |L\rangle = |1\rangle$) qubits according to
\begin{eqnarray}
\lefteqn{c_{HR}|HR\rangle + c_{HL}|HL\rangle + c_{VR}|VR\rangle + c_{VL}|VL\rangle \longrightarrow }
\nonumber \\
&&c_{HR}|HR\rangle + c_{HL}|HL\rangle + c_{VR}|VL\rangle + c_{VL}|VR\rangle,
\end{eqnarray}
will transform the four-mode coherent-state input with eigenvalues 
\begin{eqnarray}
\lefteqn{\hspace*{-.5in}
\left[\begin{array}{cccc} \langle\hat{a}_{HR}\rangle & \langle\hat{a}_{HL}\rangle & \langle\hat{a}_{VR}\rangle & \langle\hat{a}_{VL}\rangle \end{array}\right] = }\nonumber \\
&&\hspace*{.25in}\left[\begin{array}{cccc} \alpha_{HR} & \alpha_{HL} & \alpha_{VR} & \alpha_{VL}\end{array}\right],
\end{eqnarray} 
into a four-mode coherent-state output with eigenvalues
\begin{eqnarray}
\lefteqn{\hspace*{-.5in}\left[\begin{array}{cccc} \langle\hat{a}_{HR}\rangle & \langle\hat{a}_{HL}\rangle & \langle\hat{a}_{VR}\rangle & \langle\hat{a}_{VL}\rangle \end{array}\right] = }\nonumber \\
&&\hspace*{.25in}
\left[\begin{array}{cccc} \alpha_{HR} & \alpha_{HL} & \alpha_{VL} & \alpha_{VR}\end{array}\right],
\end{eqnarray} 
where the $\hat{a}$'s are annihilation operators for modes labeled by their polarization and beam positions.  It follows that the coherent-state $P_E$ and $I_B$ calculations mimic the qubit derivations that we presented in Sec.~III, with coherent-state inner products taking the place of qubit-state inner products.  At low average photon number, these coherent-state results reduce to the qubit expressions for events which give rise to clicks in the photodetectors shown in Figs.~6--9.

Finally, a word about propagation loss, detector inefficiencies, and extraneous counts from dark current or background light is in order.  All of these non-idealities actually help our Eve, in that they lead to a non-zero quantum bit error rate between Alice and Bob in the absence of the FPB attack.  If Eve's $P_E$ value is set below that baseline error rate, then her presence should be undetectable.  

\vspace*{.2in}
\begin{acknowledgments}
The authors acknowledge useful technical discussions with Howard Brandt, Jonathan Smith and Stewart Personick.  This work was supported by the Department of Defense Multidisciplinary University Research Initiative program under Army Research Office grant DAAD-19-00-1-0177 and by MIT Lincoln Laboratory. 
\end{acknowledgments} 

\appendix*
\section{QND Measurement}
Here we show that it is possible, in principle, to use cross-phase modulation between a strong coherent-state probe beam and an arbitrarily-polarized signal beam to make a QND measurement of the signal beam's total photon number.  Let $\{\hat{a}_H, \hat{a}_V, \hat{a}_P\}$ be the annihilation operators of the horizontal and vertical polarizations of the signal beam and the (single-polarization) probe beam, respectively at the input to a cross-phase modulation interaction.  We shall take that interaction to transform these annihilation operators according to the following commutator-preserving unitary operation,
\begin{eqnarray}
\hat{a}_H &\longrightarrow& \hat{a}'_H \equiv \exp(i\kappa \hat{a}_P^\dagger\hat{a}_P)\hat{a}_H \\ 
\hat{a}_V &\longrightarrow& \hat{a}'_V \equiv \exp(i\kappa \hat{a}_P^\dagger\hat{a}_P)\hat{a}_V\\
\hat{a}_P &\longrightarrow&\hat{a}'_P \equiv  \exp[i\kappa(\hat{a}_H^\dagger\hat{a}_H + \hat{a}_V^\dagger\hat{a}_V)]\hat{a}_P,  
\end{eqnarray}
where $0 < \kappa \ll 1$ is the cross-phase modulation coupling coefficient.  When the probe beam is in a strong coherent state, $|\sqrt{N}_P\rangle$ with $N_P\gg 1/\kappa^2$, the total photon number in the signal beam can be inferred from a homodyne-detection measurement of the appropriate probe quadrature.  In particular,  the state of $\hat{a}'_P$ will be $|\sqrt{N}_P\rangle$ when the signal beam's total photon number is zero, and its state will be $|(1+i\kappa)\sqrt{N}_P\rangle$ when the signal beam's total photon number is one, where $\kappa \ll 1$ has been employed.  Homodyne detection of the $\hat{a}'_{P2} \equiv {\rm Im}(\hat{a}'_P)$ quadrature thus yields a classical random-variable outcome $\alpha'_{P2}$ that is Gaussian distributed with mean zero and variance 1/4, in the absence of a signal-beam photon, and Gaussian distributed with mean $\kappa\sqrt{N}_P$ and variance 1/4 in the presence of a signal-beam photon.  Note that these conditional distributions are independent of the polarization state of the signal-beam photon when it is present.  Using the decision rule, ``declare signal-beam photon present if and only if $\alpha'_{P2} > \kappa\sqrt{N}_P/2$,'' it is easily shown that the QND error probability is bounded above by $\exp(-\kappa^2 N_P/2)/2 \ll 1$.  

The preceding polarization independent, low error probability QND detection of the signal beam's total photon number does \em not\/\rm\ disturb the polarization state of that beam.  This is so because the probe imposes the same nonlinear phase shift on both the $H$ and $V$ polarizations of the signal beam.  Hence,  
if the signal-beam input is in the arbitrarily-polarized single-photon state, 
\begin{equation}
|\psi_S\rangle = c_H |1\rangle_{H}|0\rangle_{V} + c_V |0\rangle_{H}|1\rangle_{V},
\vspace*{.075in}
\end{equation}
where $|c_H|^2 + |c_V|^2 = 1$, then, except for a physically unimportant absolute phase, the signal-beam output will also be in the state $|\psi_S\rangle$.

\end{document}